\documentclass[twocolumn,pra,aps,superscriptaddress,showpacs]{revtex4}
\usepackage{amsmath}
\usepackage{graphicx}

\begin{document}
\title{Fractional windings of the spinor condensates on a ring}
\author{Yong-Kai Liu}
\affiliation{Department of Physics, Beijing Normal University, Beijing 100875, China}
\author{Shi-Jie Yang\footnote{Corresponding author: yangshijie@tsinghua.org.cn}}
\affiliation{Department of Physics, Beijing Normal University, Beijing 100875, China}
\affiliation{State Key Laboratory of Theoretical Physics, Institute of Theoretical Physics, Chinese Academy of Sciences£¬Beijing, 100190}
\begin{abstract}
We study the uniform solutions to the one-dimensional spinor Bose-Einstein condensates on a ring. These states explicitly display the associated motion of the super-current and the spin rotation, which give rise to fractional winding numbers according to the various compositions of the hyperfine states. It simultaneously yields a fractional factor to the global phase due to the spin-gauge symmetry. All fractional windings can be denoted as $nk/(m+n)$, with $nk<m+n<2F$, for arbitrary spin-$F$ Bose-Einstein condensation (BEC). Our method can be applied to explore the fractional vortices by identifying the ring as the boundary of two-dimensional (2D) spinor condensates.
\end{abstract}
\maketitle

\section{introduction}
The Bose-Einstein condensation (BEC) of spinor atoms have been realized in the optical traps which confine the atoms regardless of their spin hyperfine states\cite{Pogosov,Mizushima2}. Due to the spin degrees of freedom, there exist a variety of complicated structures such as magnetic crystallization, spin textures as well as fractional vortices are allowed\cite{MH1,MH2,Yip,Barnett,Ciobanu,MH3,Kawaguchi,Ueda}. The direction of the atom spin can change dynamically due to collisions between the atoms. The mean-field Hamiltonian of the spinor BEC is invariant under combined global gauge transformation and spin rotation $U(1)_{\textrm{phase}}\times SO(3)_{\textrm{spin}}$\cite{Yip}. This spin-gauge symmetry reveals an interplay between superfluidity and magnetism. A spatial variation of phase is always associated an internal rotation in the spin space, giving rise to a global phase in the condensate. This phase can be fractional times of $2\pi$ which leads to a fractional vortex\cite{Kobayashi,Ho,Ohmi,Marzlin,Ogawa,Mizushima1,Ruostekoski,Gordon,Roberto,Griesmaier,Leanhardt}.

The symmetries of the Hamiltonian are usually not completely broken in the ground state. According to the homotopy theory, the broken symmetry in the ground state specifies which type of the topological excitations, such as vortex, monopole, skyrmion, et al, can be hosted in that phase.\cite{Kawaguchi,Ueda}. The set of operations under which the order parameter remains invariant constitutes an isotropy group to characterize the residual symmetry of the phase. Its conjugate class provides a label to classify the individual states.

In this paper, we address the classification of fractional windings of the spin for an arbitrary spin-$F$ ($F=1,2,3$ as example) BEC on a uniform ring. Non-uniform solutions are not appropriate to the present situation\cite{Zhang}. We start from the simple plane wave solutions in which only two of the hyperfine states are nonzero. We pay our attentions to the symmetry and windings of the spin and the global phase around a circle on the ring. It shows that the states can be specified by a pair of quantum numbers ($p$,$q$), with the first number $p$ ($\times 2\pi$) specifying the global phase change around the ring and the second number $q$ ($\times 2\pi$) the spin rotation. In general, $p$ and $q$ are fractional numbers. The corresponding state is said to have a fractional winding number. All fractional windings in the spin-$F$ BEC can be denoted by $nk/(m+n)$, with the constriction of $nk<m+n<2F$. Furthermore, if we identify the boundary of a two-dimensional (2D) spinor BEC to the 1D ring with large perimeter, then the ring state reflects possible topological defects residing in the 2D system. It may provide a way to discern or create fractional vortices in the spinor BEC.

\section{spin-1 condensate}
The mean-field Hamiltonian of the spinor BEC is divided into two parts,
\begin{equation}
\hat{H}=\hat{H}_0+\hat{H}_{s},
\end{equation}
where the noninteracting part of the Hamiltonian is
\begin{equation}
\hat{H}_0=\int d{\bf r} \sum_m\hat{\psi}_m^\dag [-\frac{\hbar^2}{2M}\nabla^2+V({\bf r})]\hat{\psi}_m
\end{equation}
with $V({\bf r})$ the external potential. The interaction part of the Hamiltonian $\hat{H}_{s}$ is dependent of the spin of the BEC. For $F=1$ it reads\cite{Ueda}
\begin{equation}
 \hat{H}_{s}=\frac{1}{2}\int d{\bf r} [c_0:\hat{n}^2:+c_1:{\bf\hat{f}}^2:],
\end{equation}
where $\bf\hat f$ are the three spin matrices and $\hat{n}$ is the density operator. The interaction strength $c_0$ and $c_1$ are related to the $s$-wave scattering length of total spin $F=0$ and $F=2$ channels, respectively.

The dynamical motion of the condensate is governed by the Heisenberg equation, which is explicitly written as the coupled Gross-Pitaevskii equations (GPEs),
\begin{equation}
i\hbar\frac{\partial\psi_m}{\partial t}=-\frac{\hbar^2}{2M}\nabla^2\psi_m+
c_0n\psi_m+c_1\sum_{n=-1}^1{\bf F}\cdot {\bf\hat f}_{mn}\psi_n,\label{a}
\end{equation}
where $m=0,\pm 1$ indicate the hyperfine states and the spin polarization ${\bf F}=\sum_m\langle\psi_m| {\bf\hat f}|\psi_m\rangle$. The wave function should satisfy the periodic boundary conditions $\psi_m(L)=\psi_m(0)$ on the ring. The stationary equations are obtained by substituting $\psi_m(x,t)=\psi_m(x)e^{-i\mu t/\hbar}$ into Eqs.(\ref{a}),
\begin{widetext}
\begin{eqnarray}
\mu\psi_1&=&[-\frac{1}{2}\partial_x^2+(c_0+c_2)(\vert\psi_1\vert^2+\vert\psi_0\vert^2)+(c_0-c_2)
\vert\psi_{-1}\vert^2]\psi_1+c_2\psi_0^2\psi_{-1}^* \nonumber\\
\mu\psi_0&=&[-\frac{1}{2}\partial_x^2+(c_0+c_2)(\vert\psi_1\vert^2+\vert\psi_{-1}\vert^2)+c_0
\vert\psi_0\vert^2]\psi_0+2c_2\psi_0^*\psi_1\psi_{-1} \nonumber\\
\mu\psi_{-1}&=&[-\frac{1}{2}\partial_x^2+(c_0+c_2)(\vert\psi_{-1}\vert^2+\vert\psi_0\vert^2)+(c_0-c_2)
\vert\psi_{1}\vert^2]\psi_{-1}+c_2\psi_0^2\psi_1^* \label{F1}
\end{eqnarray}
\end{widetext}

We consider the unform plane wave solution with $\psi_0=0$ as
\begin{equation}
\psi=\left(\begin{array}{c}Ae^{i\theta}\\0\\D
\end{array}\right)\label{wavefun1}
\end{equation}
where $\theta=2\pi x/L$ with $L$ the perimeter of the ring. Here "uniform" indicates that the profile of density is uniform on the ring. $A$ and $D$ are constants which can be directly calculated from Eqs.(\ref{F1}),
\begin{eqnarray}
A^2&=&\frac{\mu}{2c_0}-\frac{\pi^2}{L^2}(\frac{1}{2c_0}+\frac{1}{2c_2}),\nonumber\\
D^2&=&\frac{\mu}{2c_0}-\frac{\pi^2}{L^2}(\frac{1}{2c_0}-\frac{1}{2c_2}).
\end{eqnarray}
In the limit of $L\rightarrow\infty$, the corresponding ground state $\psi^{\textrm{ground}}=(A,0,D)^T$ becomes a polar state $\psi^{\textrm{polar}}$ (with $A=D=1/\sqrt{2}$) which has a continuous $SO(2)$ symmetry and a discrete $C_{2}$ symmetry perpendicular to the $SO(2)$ symmetrical axis. The underlying reason is that $L\rightarrow\infty$ implies that the kinetic energy is ignorable. For the finite ring sample ($L$ is finite), $A\neq D$ indicates only a discrete $C_{2z}$ symmetry in (\ref{wavefun1}) since a super-current is excited along the ring. It is called the $C_2$ state\cite{Kobayashi}. One notes that the mass circulation $\oint_L\upsilon^{(\textrm{mass})}dx=\frac{h}{M}\cdot\frac{A^2}{A^2+D^2}$ is not simply quantized.

As the full symmetry group of the Hamiltonian is $U(1)_{\textrm{phase}}\times SO(3)_{\textrm{spin}}$, the mass flow of one hyperfine state will simultaneously induce a rotation in the spin space. It equivalently yields a global flow of the condensate. The order parameter is described by $\psi=\sqrt{n}\zeta=\sqrt{n}e^{i\phi}U(\alpha,\beta,\gamma)\zeta_0$, where $\phi$ is the global gauge and  $U(\alpha,\beta,\gamma)=e^{-if_z\alpha}e^{-if_y\beta}e^{-if_z\gamma}$ is the rotational matrix. $\zeta_0$ is the reference order parameter. By make a mapping $\alpha=-\phi=-\theta/2$ from physical space into the order parameter space, we decompose the state (\ref{wavefun1}) as $\psi=e^{i\theta/2}U(-\theta/2,0,0)(A,0,D)^T$, where the 1D path is parameterized by the angular coordinate $\theta$. In the state (\ref{wavefun1}), $\theta$ increases $2\pi$ as one goes one circle along the ring.

Figure 1 display the density distribution in the spherical-harmonic representation as\cite{Kawaguchi,Ueda}.
\begin{equation}
\Psi({\bf r},\hat{s})=\sum_m\psi_m({\bf r})Y_{Fm}(\hat{s}).
\end{equation}
The parameter are chosen as $a_0=114 a_B$, $a_2=103 a_B$, $\mu=52.45$ . It is seen that the the state exhibits the $C_{2z}$ symmetry. The ${\bf d}$ vector describes the spin rotation when one goes along the ring. In this case, vector ${\bf d}$ rotates an angle of $-\pi$ around the ring.

Since $\psi(L)=e^{i\pi}e^{i{\hat f}_z\pi}\psi(0)$, we get the global phase $\phi=\frac{1}{2}\times 2\pi$ and the associated spin-rotation angle $\alpha=-\frac{1}{2}\times 2\pi$ around the $z$-axis. Consequently, the state (\ref{wavefun1}) can be represented by a pair of quantum numbers $(\frac{1}{2},\frac{1}{2})$. In general, an arbitrary current-carrying state on the ring can be specified by a pair of quantum numbers $(p,q)$ with the $p$ specifying the change of the global phase and $q$ the change of the spin-rotating angle\cite{Ueda}. We observe that $p$ and $q$ are related to the mass current $\upsilon^{mass}=\frac{\hbar}{M}(\nabla\phi-|f|\nabla\alpha)$ by $p=\frac{1}{2\pi}\oint\nabla\phi\cdot d \ell$ and $q=\frac{1}{2\pi}\oint\nabla\alpha\cdot d\ell$. Moreover, $p$ and $q$ are usually fractional numbers and the state is called to have fractional windings. Since the 1D ring can be identified to the boundary of the 2D system, the state $\psi$ reflects a fractional vortex with winding numbers $(\frac{1}{2},\frac{1}{2})$ residing in the 2D spinor BEC, due to the topological invariance. The vortex core may be filled with normal phases\cite{Kobayashi}.

\begin{figure}[t]
\begin{center}
\includegraphics*[width=8cm]{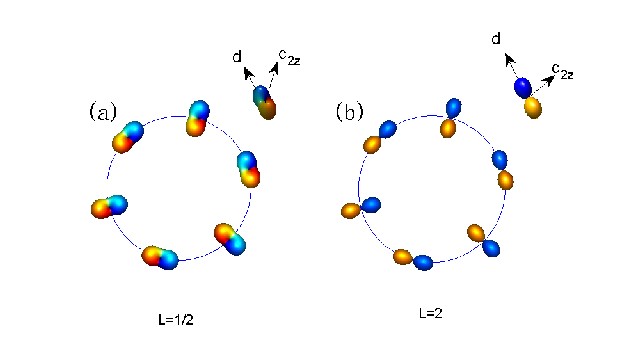}
\end{center}
\caption{The spherical harmonic representation of the ($\frac{1}{2},\frac{1}{2}$) state of the spin-1 BEC. (a) $L=1/2$. (b) $L=2$.}
\end{figure}

\section{spin-2 condensate}
In this section we address the classification of fractional winding states of $F=2$ BEC on a ring. The interaction part of the Hamiltonian is\cite{M}
\begin{equation}
\hat{H}_{s}=\frac{1}{2}\int d{\bf r} [c_0:\hat{n}^2:+c_1:\hat{\bf f}^2:+c_2\hat{A}_{00}^\dag\hat{A}_{00}],
\end{equation}
where $\hat{A}_{00}=\frac{1}{\sqrt{5}}(2\hat{\psi}_{2}\hat{\psi}_{-2}-2\hat{\psi}_{1}\hat{\psi}_{-1}+\hat{\psi}_{0}^2)$. The stationary GPEs are
\begin{widetext}
\begin{eqnarray}
\mu\psi_{\pm2}&=&[-\frac{1}{2}\partial_x^2+c_0n\pm2 c_1F_z]\psi_{\pm2}+c_1F_{\mp}\psi_{\pm1}+\frac{c_2}{\sqrt{5}}A_{00}\psi_{\mp2}^*,\nonumber\\
\mu\psi_{\pm1}&=&[-\frac{1}{2}\partial_x^2+c_0n\pm c_1F_z]\psi_{\pm1}+c_1(\frac{\sqrt{6}}{2}F_\mp\psi_0+F_\pm\psi_{\pm2})-\frac{c_2}{\sqrt{5}}A_{00}\psi_{\mp1}^*,\nonumber\\
\mu\psi_0&=&[-\frac{1}{2}\partial_x^2+c_0n]\psi_0+\frac{\sqrt{6}}{2}c_1 (F_+\psi_1+F_-\psi_{-1})+\frac{c_2}{\sqrt{5}}A_{00}\psi_0^*,\label{F2}
\end{eqnarray}
\end{widetext}
where $F_+=F_-^*=2(\psi_{2}^*\psi_{1}+\psi_{-1}^*\psi_{-2})+\sqrt{6}(\psi_{1}^*\psi_{0}+\psi_{0}^*\psi_{-1})$, $F_z=2(|\psi_{2}|^2-|\psi_{-2}|^2)+|\psi_{1}|^2-|\psi_{-1}|^2$ and $A_{00}=\frac{1}{\sqrt{5}}(2\psi_{2}\psi_{-2}-2\psi_{1}\psi_{-1}+\psi_{0}^2)$.

We consider the following forms of uniform solution in which only two of the hyperfine states are nonzero,
\begin{widetext}
\begin{eqnarray}
\psi_1=\left(\begin{array}{c}Ae^{i\theta}\\0\\0\\D\\0\end{array}\right),\hspace{2mm}
\psi_2=\left(\begin{array}{c}Ae^{i\theta}\\0\\0\\0\\D\end{array}\right),\hspace{2mm}
\psi_3=\left(\begin{array}{c}0\\Ae^{i\theta}\\0\\D\\0\end{array}\right),\hspace{2mm}
\psi_4=\left(\begin{array}{c}Ae^{i\theta}\\0\\D\\0\\Ae^{-i\theta}
\end{array}\right).
\label{wavefun2}
\end{eqnarray}
\end{widetext}
\begin{figure}[t]
\begin{center}

\includegraphics*[width=9cm]{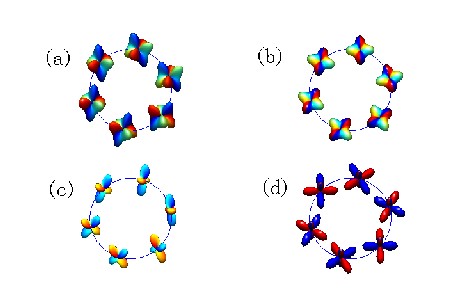}
\end{center}
\caption{The order parameter of the spin-2 BEC for states specified as (a) $\psi_1\rightarrow(\frac{1}{3},\frac{1}{3})$, (b) $\psi_2\rightarrow(\frac{1}{2},\frac{1}{4})$, (c) $\psi_3\rightarrow(\frac{1}{2},\frac{1}{2})$, and (d) $\psi_4\rightarrow(0,\frac{1}{2})$.}
\end{figure}
It is straightforward to calculate the constants $A$ and $D$ in each solution by substituting it into Eqs.(\ref{F2}). The symmetry of the order parameters are analyzed by the same way as in the $F=1$ BEC in the previous section. The ground state of $\psi^{\textrm{ground}}_1=(A,0,0,D,0)^T$ is compared with cyclic state $(\frac{1}{\sqrt{3}},0,0,\frac{\sqrt{2}}{\sqrt{3}},0)^T$ which has the tetrahedral symmetry $T$\cite{Kawaguchi}. However, the symmetry of the $\psi_1$ state degenerates into the $C_{3z}$ symmetry by the same reason as in the spin-1 BEC that a super-current is excited. As $\theta$ increases from $0$ to $2\pi$ around the ring, the spin rotates an angle of $-2\pi/3$ associated to a global phase of $2\pi/3$ to yield $\psi_1(L)=e^{2\pi/3i}e^{i{\hat f}_z 2\pi/3}\psi_1(0)$. Hence the $\psi_1$ state has the $(\frac{1}{3},\frac{1}{3})$ fractional winding numbers with the $C_{3z}$ symmetry, as shown in Fig.2(a).

The ground state of $\psi_2^{\textrm{ground}}=(A,0,0,0,D)^T$ is compared to the biaxial nematic state $(\frac{1}{\sqrt{2}},0,0,0,\frac{1}{\sqrt{2}})^T$ which has the $D_4$ symmetry\cite{Kawaguchi}. Similarly, the symmetry of $\psi_2$ has only the $C_{4z}$ symmetry. As $\theta$ increase from $0$ to $2\pi$, $\psi_2(L)=e^{i\pi}e^{i{\hat f}_z \pi/2\psi_2}$. Hence the state $\psi_2$ has
 fractional winding numbers $(\frac{1}{2},\frac{1}{4})$, as is shown in Fig.2(b).

By the same way, we find that the $\psi_3$ state has the $C_{2z}$ symmetry and can be specified by the winding numbers of $(\frac{1}{2},\frac{1}{2})$. The ground state of $\psi_4^{\textrm{ground}}=(A,0,D,0,-A)^T$ is compared to the cyclic state $(\frac{1}{2},0,\frac{\sqrt{2}}{2},0,-\frac{1}{2})^T$. However, the $\psi_4$ state has only the $C_{2z}$ symmetry due to the super-current in the ring. It is specified by the winding numbers of $(0,\frac{1}{2})$.

As to the argument in the $F=1$ BEC, from the states $\psi_1$, $\psi_2$, $\psi_3$, and $\psi_4$ we conclude that the 2D $F=2$ BEC may host $(\frac{1}{3},\frac{1}{3})$, $(\frac{1}{2},\frac{1}{4})$, $(\frac{1}{2},\frac{1}{2})$, and $(0,\frac{1}{2})$ fractional vortices, respectively\cite{Ueda}.

\section{spin-3 condensate}
The interaction part of the $F=3$ Hamiltonian is\cite{Kawaguchi,Ueda},
\begin{equation}
\hat{H}_{s}=\frac{1}{2}\int d{\bf r} [c_0:\hat{n}^2:+c_1:\hat{\bf f}^2:+c_2\hat{A}_{00}^\dag\hat{A}_{00}
+c_3\sum_M\hat{A}_{2M}^\dag\hat{A}_{2M}],
\end{equation}
where $\hat{A}_{00}=\frac{1}{\sqrt{7}}(2\hat{\psi}_{3}\hat{\psi}_{-3}-2\hat{\psi}_{2}\hat{\psi}_{-2}+2\hat{\psi}_{1}\hat{\psi}_{-1}-\hat{\psi}_{0}^2)$, $\hat{A}_{20}=\frac{1}{\sqrt{7}}(\frac{5}{\sqrt{3}}\hat{\psi}_{3}\hat{\psi}_{-3}-\sqrt{3}\hat{\psi}_{1}\hat{\psi}_{-1}+\sqrt{\frac{4}{3}}\hat{\psi}_{0}^2)$,
$\hat{A}_{2\pm1}=\frac{1}{\sqrt{7}}(\frac{5}{\sqrt{3}}\hat{\psi}_{\pm3}\hat{\psi}_{\mp2}-\sqrt{5}\hat{\psi}_{\pm2}\hat{\psi}_{\mp1}+\sqrt{\frac{2}{3}}\hat{\psi}_{\pm1}\hat{\psi}_{0})$, and $\hat{A}_{2\pm2}=\frac{1}{\sqrt{7}}(\sqrt{\frac{10}{3}}\hat{\psi}_{\pm3}\hat{\psi}_{\mp1}-\sqrt{\frac{20}{3}}\hat{\psi}_{\pm2}\hat{\psi}_{0}+\sqrt{2}\hat{\psi}_{\pm1}^2)$. The relevant physical parameters can be found in Ref.[\onlinecite{Kawaguchi}]. The stationary GPEs are written as,
\begin{widetext}
\begin{eqnarray}
\mu\psi_{\pm3}&=& [-\frac{1}{2}\partial_x^2+c_0n\pm3c_1F_z]\psi_{\pm3}+\frac{\sqrt{6}}{2}c_1F_{\mp}\psi_{\pm2} +\frac{c_2}{\sqrt{7}}A_{00}\psi_{\mp3}^*+\frac{c_3}{2\sqrt{21}}[5 A_{20}\psi_{\mp3}^*+
5 A_{2\pm1}\psi_{\mp2}^*+\sqrt{10}A_{2\pm2}\psi_{\mp1}^*],\nonumber\\
\mu\psi_{\pm2}&=&[-\frac{1}{2}\partial_x^2+c_0n\pm2c_1F_z]\psi_{\pm2}+\frac{c_1}{2}[\sqrt{10}F_{\mp}\psi_{\pm1}
+\sqrt{6}F_\pm\psi_{\pm3}]-\frac{c_2}{\sqrt{7}}A_{00}\psi_{\mp2}^*\nonumber\\
&&+\frac{c_3}{2\sqrt{21}}[5 A_{2\mp1}\psi_{\mp3}^*-\sqrt{15}A_{2\pm1}\psi_{\mp1}^*-
\sqrt{20}A_{2\pm2}\psi_{0}^*],\nonumber\\
\mu\psi_{\pm1}&=&[-\frac{1}{2}\partial_x^2+c_0n\pm c_1F_z]\psi_{\pm1}+\frac{c_1}{2}[\sqrt{12}F_{\mp}\psi_{0}+
\sqrt{10} F_\pm\psi_{\pm2}]+\frac{c_2}{\sqrt{7}}A_{00}\psi_{\mp1}^*\nonumber\\
&&+\frac{c_3}{2\sqrt{7}} [\frac{\sqrt{10}}{3}A_{2\mp2}\psi_{\mp3}^*-\sqrt{5}A_{2\mp1}\psi_{\mp2}^*-\sqrt{3}A_{20}\psi_{\mp1}^*
+{2\sqrt{2}A_{2\pm2}\psi_{\pm1}^*}+\frac{\sqrt{2}}{3}A_{2\pm1}\psi_0^*],\nonumber\\
\mu\psi_{0}&=&[-\frac{1}{2}\partial_x^2+c_0n]\psi_{0}+\sqrt{3}c_1[F_-\psi_{-1}+F_+\psi_{1}] -\frac{1}{\sqrt{7}}c_2A_{00}\psi_0^*\nonumber\\
&&+\frac{c_3}{2\sqrt{21}}[4 A_{20}\psi_0^*+\sqrt{2} A_{21}\psi_1^*+
\sqrt{2} A_{2-1}\psi_{-1}^*
-\sqrt{20} A_{22}\psi_2^*-\sqrt{20} A_{22}\psi_{-2}^*].\label{F3}
\end{eqnarray}
\end{widetext}

We consider the following combinations of uniform plane wave solutions,
\begin{widetext}
\begin{eqnarray}
\psi_1=\left(\begin{array}{c}Ae^{i\theta}\\0\\0\\0\\D\\0\\0\end{array}\right),\hspace{2mm}
\psi_2=\left(\begin{array}{c}Ae^{i\theta}\\0\\0\\0\\0\\D\\0\end{array}\right),\hspace{3mm}
\psi_3=\left(\begin{array}{c}Ae^{i\theta}\\0\\0\\0\\0\\0\\D\end{array}\right),\hspace{3mm}
\psi_4=\left(\begin{array}{c}0\\Ae^{i\theta}\\0\\0\\D\\0\\0\end{array}\right),\hspace{2mm}
\psi_5=\left(\begin{array}{c}0\\Ae^{i\theta}\\0\\0\\0\\D\\0\end{array}\right).
\label{wavefun3}
\end{eqnarray}
\end{widetext}
The constants $A$ and $D$ in each state can be calculated directly by substituting the state into the GPEs (\ref{F3}).
\begin{figure}
\begin{center}
\includegraphics*[width=9cm]{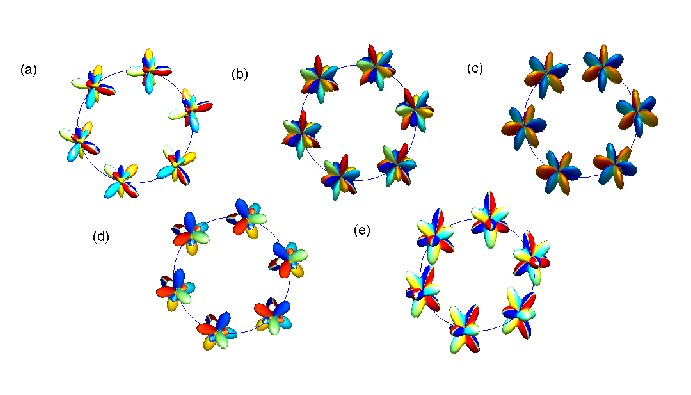}
\end{center}
\caption{The order parameter of the spin-3 BEC for states specified as (a) $\psi_1\rightarrow(\frac{1}{4},\frac{1}{4})$, (b) $\psi_2\rightarrow(\frac{2}{5},\frac{1}{5})$, (c) $\psi_3\rightarrow(\frac{1}{2},\frac{1}{6})$, (d) $\psi_4\rightarrow(\frac{1}{3},\frac{1}{3})$, and (e) $\psi_5\rightarrow(\frac{1}{2},\frac{1}{4})$.}
\end{figure}

The symmetry of the order parameters can be analyzed as in the previous sections. The phase diagram of the spin-3 BEC has been investigated in Ref.[\onlinecite{Kawaguchi}]. The state $\psi_1$ has the $C_{4z}$ symmetry. As $\theta$ increases from $0$ to $2\pi$ around the ring, the spin rotates an angle of $-\pi/4$ associated to a global phase of $\pi/4$ which yields $\psi_1(L)=e^{\pi/2i}e^{i{\hat f}_z \pi/2}\psi_1(0)$. Hence the $\psi_1$ state has the $C_{4z}$ symmetry with $(\frac{1}{4},\frac{1}{4})$ winding numbers, as shown in Fig.3(a). The state $\psi_2$ has the discrete $C_{5z}$ symmetry. When one goes around the ring, the spin rotates an angle of $-2\pi/5$ associated to a global phase of $4\pi/5$ which yields $\psi_2(L)=e^{4\pi/5i}e^{i{\hat f}_z 2\pi/5}\psi_1(0)$. Hence the state $\psi_2$ is specified by a pair of fractional winding numbers $(\frac{2}{5},\frac{1}{5})$. Similarly, the state $\psi_3$ has the $C_{6z}$ symmetry and is specified by the winding numbers of $(\frac{1}{2},\frac{1}{6})$. The state $\psi_4$ has the $C_{3z}$ symmetry and is specified by $(\frac{1}{3},\frac{1}{3})$. The state $\psi_5$ has the $C_{4z}$ symmetry and is specified by $(\frac{1}{2},\frac{1}{4})$.

\begin{figure}[h]
\begin{center}
\includegraphics*[width=9cm]{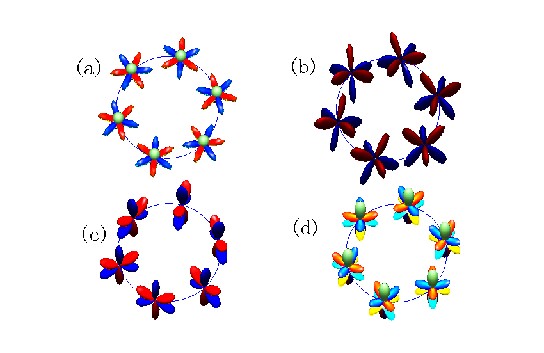}
\end{center}
\caption{The order parameter of the spin-3 BEC for states specified as (a) $\psi_6\rightarrow(0,\frac{1}{3})$, (b) $\psi_7\rightarrow(0,\frac{1}{3})$, (c) $\psi_8\rightarrow(0,\frac{1}{2})$, and (d) $\psi_9\rightarrow(0,\frac{1}{2})$.}
\end{figure}

There is another class of combinations of the uniform state in which $\psi_0\neq 0$. They are explicitly as follows,
\begin{widetext}
\begin{eqnarray}
\psi_6=\left(\begin{array}{c}Ae^{i\theta}\\0\\0\\D\\0\\0\\Ae^{-i\theta}\end{array}\right),\hspace{2mm}
\psi_7=\left(\begin{array}{c}Ae^{i\theta}\\0\\0\\D\\0\\0\\-Ae^{-i\theta}\end{array}\right),\hspace{2mm}
\psi_8=\left(\begin{array}{c}0\\Ae^{i\theta}\\0\\D\\0\\Ae^{-i\theta}\\0\end{array}\right),\hspace{2mm}
\psi_9=\left(\begin{array}{c}0\\Ae^{i\theta}\\0\\D\\0\\-Ae^{-i\theta}\\0\end{array}\right).
\label{wavefun4}
\end{eqnarray}
\end{widetext}
The symmetry and winding numbers are analyzed by the same way. The state $\psi_6$ has the $D_{3z}$ symmetry and is specified by $(0,\frac{1}{3})$. The state $\psi_7$ has the $C_{3z}$ symmetry and is also specified by $(0,\frac{1}{3})$. The state $\psi_8$ has the $D_2$ symmetry and is specified by $(0,\frac{1}{2})$. Finally, the state $\psi_9$ has the $C_{2z}$ symmetry and is specified by $(0,\frac{1}{2})$. This classification of states can be applied to the classification of fractional vortices in the 2D spin-3 BEC. From the states $\psi_1\sim \psi_9$, we conclude that the 2D $F=3$ BEC may topologically host the $\frac{1}{5}$, $\frac{1}{4}$, $\frac{1}{3}$, and $\frac{1}{2}$ fractional vortices.

\section{Conclusion}
We have studied the uniform solutions to arbitrary spin-$F$ BEC on a ring. These states consist of various combinations of the hyperfine states. The symmetry of the states are analyzed. We demonstrated that the states can be specified by a pair of quantum numbers which describe the global phase and the windings of the spin. According to our method, the possible fractional windings in spin-$F$ BEC can denoted by $nk/(m+n)$, with the constriction of $nk<m+n<2F$. For example, in $F=1$ BEC we have $1/2$ fractional winding but have no $1/3$ winding; in $F=2$ BEC we have $1/2$ and $1/3$ fractional windings but have no $1/4$ and $1/5$ windings; in $F=3$ BEC we have $1/2$, $1/3$, $1/4$ and $1/5$ windings but have no $1/6$ and $1/7$ windings, and so on. Our method can help us to discern or create the fractional vortices in the 2D spinor BEC.

This work is supported by the funds from the Ministry of Science and Technology of China under Grant No. 2012CB821403.

\end{document}